\documentclass[smallextended]{svjour3}       
\smartqed  

\usepackage{graphicx}

\begin{document}

\title{Angular diameter distances reconsidered in the Newman \& Penrose formalism}

\author{Thomas P. Kling         \and
        Aly Aly
}


\institute{T. Kling \at
              Dept. of Physics, Bridgewater State University,
Bridgewater, MA 02325 \\
              Tel.: +1-508-531-2895\\
              Fax: +1-508-531-1785\\
              \email{tkling@bridgew.edu}           
           \and
           A. Aly \at
              Dept. of Physics, Bridgewater State University,
Bridgewater, MA 02325
}

\date{Received: date / Accepted: date}

\maketitle

\begin{abstract}

\noindent Using the Newman and Penrose spin coefficient (NP) formalism, we
provide a derivation of the Dyer-Roeder equation for the angular diameter distance in cosmological
space-times.  We show that the geodesic deviation equation written in NP formalism is precisely the Dyer-Roeder equation for a general Friedman-Robertson-Walker (FRW) space-time, and then we examine the angular diameter distance to redshift relation in the case that a flat FRW metric is perturbed by a gravitational potential.  We examine the perturbation in the case that the gravitational potential exhibits the properties of a thin gravitational lens, demonstrating how the weak lensing shear and convergence act as source terms for the perturbed Dyer-Roeder equation.

\end{abstract}

\PACS{95.30.-k, 95.30.Sf, 04.90.+e}

\maketitle


\section{Introduction} \label{intro:sec}

The Newman-Penrose (NP) spin coefficient formalism has played a crucial role in the history of general relativity \cite{NP}.  It has been most often utilized in the context of matter free space-times, either in terms of providing a path to deriving solutions to the Einstein Field equations, or in terms of providing a framework to study gravitational radiation.

Because the NP formalism is based on a series of null vectors, it provides a natural framework to examine the properties of space-times through an examination of pencils of light rays.  Several recent papers have worked with the  NP formalism to study gravitational lensing \cite{bianchini}, \cite{kc}.  These studies examined the relations between the observable and directly derived quantities of interest in weak gravitational lensing and components of the Ricci and Weyl tensors in the NP formalism.  In those papers, the curvature tensors were associated with matter density perturbations on flat space-times where the matter density represented super-clusters of galaxies as one would see in gravitational lensing.  The justification for using a flat background space-time was that the Friedman-Robertson-Walker (FRW) space-times are conformally flat, and that a conformal transformation of the metric leaves light rays unchanged.

In this paper, we examine the propagation of bundles of light rays through perturbed FRW space-times.  We are primarily interested in understanding how the angular-diameter distance arises as a function of the redshift in a general FRW cosmology that is perturbed by a gravitational potential.  The differential equation relating the angular-diameter distance to the redshift in a FRW cosmology is known as the Dyer-Roeder equation. The original paper of Dyer and Roeder was historically important in developing an understanding of the impact that cosmological expansion has on distance scales\cite{dyer}.  The Dyer-Roeder equation has been re-examined to shed light on properties of quintessence cosmologies \cite{giovi}, \cite{lewis}.  Recent studies also examine the Dyer-Roeder equation in cosmological space-times, applying their results to supernovae observations in \cite{clarkson1}, and distances to the CMB in \cite{clarkson2} and \cite{kaiser}, in particular taking into account gravitational lensing \cite{bonvin}.

In the first part of this paper, we show that the geodesic deviation equation, written in the NP formalism directly leads to the Dyer-Roeder equation in an unperturbed FRW cosmological space-time in a clear manner.  This result requires us to consider the form of the NP null tetrad in cosmological setting.  Without taking into account the cosmological impact on the form of the null tetrad, one can not derive this fundamental relation.

Our method of deriving the Dyer-Roeder equation is similar to the derivation based on the Sachs equations in \cite{perlick} or \cite{clarkson1}, but it has the advantage that we can easily extend it to derive a perturbed Dyer-Roeder equation, where the perturbation is due to an added gravitational potential.  Our new derivation of the perturbed Dyer-Roeder equation connects seamlessly to the natural quantities in weak gravitational lensing without introducing the optical scalars as an intermediate step.  We show how the source terms in the perturbed Dyer-Roeder equation are the weak lensing shear (which is directly observable through the distortion of the shape of background galaxies) and the weak lensing convergence (which one can derive from the shear)\cite{kk}.  Finally, we examine required boundary conditions and test the variation in the angular-diameter distance across regions of the sky the size of those used in measurements of weak gravitational lensing for a single, axially symmetric lens.


\section{Dyer-Roeder Equation in Standard Cosmologies} \label{standard:sec}

In this section, we derive the Dyer-Roeder equation from the geodesic deviation equation in a FRW cosmological space-time. For generality, we do not wish to restrict ourselves to the flat cosmological scenarios, so that the metric takes the form

\begin{equation} ds^2 = -dt^2 + \frac{a^2(t)}{1 - kr^2} dr^2 + a^2(t) r^2 d\theta^2 + a^2(t) r^2 \sin^2\theta d\phi^2, \label{FRW1} \end{equation}

\noindent for $k=-1, 0, +1$ and scale factor $a(t)$ related to the redshift, $z_r$, by

\begin{equation} \frac{1}{a(t)} = 1+z_r. \label{redshift} \end{equation}

\noindent In our derivation, we will need the following components of the connection

\begin{eqnarray} \Gamma^{0}_{11} = \frac{a \dot a}{1-kr^2} &\quad\quad& \Gamma^1_{11} = \frac{kr}{1-kr^2} \nonumber \\ \Gamma^1_{01} & = & \frac{\dot a}{a}, \label{connection:general} \end{eqnarray}

\noindent for $\dot a = da/dt$.  The components of the Ricci tensor that we will require are

\begin{equation} R_{00} =\frac{-3\ddot a}{a} \quad\quad R_{11} = \frac{2k + 2\dot a^2 + a \ddot a}{1-kr^2}. \label{Ricci:general} \end{equation}

\noindent For a non-perturbed FRW metric as in Eq.~\ref{FRW1}, the Weyl tensor is zero.

\subsection{Newman-Penrose Formalism}

Of central importance to the NP spin coefficient formalism is a tetrad of null vectors associated with a light ray in the space-time.  The tangent vector to the light ray is the most important member of this tetrad, and is denoted $\ell^a$.  The other members of the null tetrad are a pair of null, complex spatial vectors $m^a$ and $\overline m^a$ which can be taken to lie in the spatial cross section of the bundle of null geodesics of which $\ell^a$ is tangent to the central ray, and a null vector $n^a$ orthogonal to $\ell^a$.  The pair of complex spatial vectors, $m^a$, will be particularly important in what follows, though not as important in the FRW metric case.  They will satisfy

\[ m^a \nabla_a \ell^b = \ell^a \nabla_a m^b = 0, \]

\noindent with the inner products $g_{ab} \ell^a m^b = 0$ and $g_{ab} m^a \overline m^b = 1$.  The null vector $n^a$ plays no significant role in the application of the NP formalism to gravitational lensing.

One can show that for the metric in Eq.~\ref{FRW1}, in $(t, r, \theta, \phi)$ coordinates, a past-directed null geodesic moving radially away from the origin has a tangent vector, $\ell^a$, takes the form

\begin{equation}  \ell^a = \left( \frac{-1}{\sqrt{2} a}, \frac{\sqrt{1-kr^2}}{\sqrt{2} a^2}, 0, 0 \right). \label{ell1} \end{equation}

\noindent Clearly $\ell^a$ is a null vector satisfying $g_{ab} \ell^a \ell^b = 0$.  The particular normalization of $\ell^a$, or the placement of the $a$ in the denominator of the time component, ensures that $\ell^a$ satisfies the condition

\[ \ell^a \nabla_a \ell^b = 0, \]

\noindent where the connection terms in Eq.~\ref{connection:general} are used.

In the NP formalism, the information coded in the Ricci and Weyl tensors is packaged in terms of a set of complex scalar functions.  The particular component of the Ricci tensor

\begin{equation} \Phi_{00} = \frac{1}{2} R_{ab} \ell^a \ell^b, \label{ricci1} \end{equation}

\noindent helps to control the geodesic deviation equation along with the $\Psi_0$ component of the Weyl tensor

\begin{equation} \Psi_0 = C_{abcd} \ell^a m^b \ell^c m^d. \label{weyl1} \end{equation}

\noindent  In the unperturbed case we are considering in this section, $\Psi_0$ is zero, but using Eq.~\ref{Ricci:general} and the tangent vector $\ell^a$, we have

\begin{equation} \Phi_{00} = -\frac{1}{2a^2} \left( \frac{\ddot a}{a} - \frac{\dot a^2}{a^2} - \frac{k}{a^2} \right). \label{ricci1a} \end{equation}

Ultimately, we want to express the Ricci tensor in terms of constant parameters that determine the cosmology and the redshift.  Using $H = \dot a /a$, one can show that

\begin{equation} \frac{\ddot a}{a} - \frac{\dot a^2}{a^2} = \frac{1}{2H} \frac{d H^2}{dt}, \label{sub1} \end{equation}

\noindent so that the first two terms in Eq.~\ref{ricci1a} are replaced with a function dependent on the Hubble parameter.  The term related to the spatial curvature in $\Phi_{00}$ can be expressed in terms of $\Omega(a)$ by using

\[ -\frac{k}{H^2 a^2} = 1 - \Omega(a), \]

\noindent where $\Omega(a)$ is the time-evolving density parameter \cite{peacock}. We will assume that we are in a matter-vacuum energy dominated universe, and we will denote components of the density parameter today as $\Omega_m$ and $\Omega_\Lambda$, with $\Omega = \Omega_m + \Omega_\Lambda$, and the value of the Hubble constant today as $H_0$.  For the Hubble parameter, we have

\begin{equation} H^2 = H_0^2 ((1+z_r)^3 \Omega_m + (1+z_r)^2 (1 - \Omega_m - \Omega_\Lambda) + \Omega_\Lambda), \label{hubble} \end{equation}

\noindent and for the time evolution of the density parameter, we have \cite{peacock}

\begin{equation} 1 - \Omega(a) = \frac{1-\Omega_m - \Omega_\Lambda}{(1-\Omega_m - \Omega_\Lambda) + a^{-2} \Omega_\Lambda + a^{-1} \Omega_m}. \label{omegatime} \end{equation}

\noindent From $a^{-1} = 1 + z_r$, we have

\begin{equation} \frac{d}{dt} = -\frac{\dot a}{a^2} \frac{d}{dz_r}. \label{timeder} \end{equation}

\noindent Putting these items together, one determines that the Ricci tensor component that controls geodesic deviation in FRW metrics Eq.~\ref{FRW1},
 is given by

\begin{equation} \Phi_{00} = \frac{3 H_0^2  \Omega_m}{4 a^5}. \label{ricci1b} \end{equation}

In the NP formalism, the geodesic deviation vector $q^a$ is decomposed into components along a tetrad of null vectors.  For a geodetic congruence \cite{penrose}, in a pencil of null geodesics emanating from a single point, neighboring rays will be abreast and the geodesic deviation vector can be decomposed into parts lying along $m^a$ and $\overline m^a$ as

\begin{equation} q^a = \xi \overline m^a + \bar \xi m^a. \label{geod1} \end{equation}

\noindent If we define $D = \ell^a \nabla_a$,

\begin{equation} \mathbf{z} = \left( \begin{array}{c} \xi \\ \bar \xi \end{array} \right) \end{equation}

\noindent as the geodesic deviation vector, and

\begin{equation} \mathbf{Q} = \left( \begin{array}{cc} \Phi_{00} & \Psi_0 \\ \overline \Psi_0 & \Phi_{00} \end{array} \right) \label{Q} \end{equation}

\noindent as the components of the Riemann and Weyl curvature tensors in the NP formalism, then the geodesic deviation equation then takes the form

\begin{equation} D^2 \mathbf{z} = - \mathbf{Q z}. \label{geod_dev}\end{equation}

\subsection{Deriving the Dyer-Roeder Equation}

Our first goal is to show that the geodesic deviation equation in the NP formalism, Eq.~\ref{geod_dev}, is the standard Dyer-Roeder equation in the case of a non-perturbed FRW metric as in Eq.~\ref{FRW1}.  With the result that the Weyl tensor is zero and the Ricci tensor is given by Eq.~\ref{ricci1b}, we see that the geodesic deviation equation is given by

\begin{equation} D^2 \xi = -\frac{3 H_0^2 \Omega_m}{4 a^5} \xi. \label{geod_dev1} \end{equation}

\noindent Because the underlying space-time is homogeneous and isotropic, all the possible geodesic deviation vectors in a pencil of light rays emanating from an observer will follow the same expansion, regardless of their orientation.  This implies that we can choose $\xi$ to be real-valued.  We will assume that the real value of $\xi$ represents the size of a stick whose tips subtend an angle $\alpha$ at the location of the observer. We posit that the angular diameter distance $d_A$ in the homogeneous and isotropic FRW metric will be given by $d_A = \xi / \alpha$.

The key in deriving the Dyer-Roeder expression from Eq.~\ref{geod_dev1} is understanding how to write $D^2 d_A$ in terms of derivatives with respect to the redshift.  We have that

\begin{equation} D d_A = \ell^a \partial_a d_A = \ell^0 \partial_t d_A + \ell^1 \partial_r d_A, \label{D1} \end{equation}

\noindent and because the spatial surfaces are homogeneous and isotropic, $\partial_r d_A = 0$.  The second derivative would then be

\[ D(D d_A) = (\ell^0 \partial_t + \ell^1 \partial_r) (\ell^0 \partial_t d_A) = (\ell^0)^2 \partial_t^2 d_A + \ell^0 (\partial_t \ell^0) \partial_t d_A. \]

\noindent With $\ell^0 = -1/(\sqrt{2} a)$ and $H = \dot a/a$, we find that

\begin{equation} D^2 d_A = \frac{1}{2a^2} \frac{d^2 d_A}{d t^2} - \frac{H}{2a^2} \frac{d d_A}{d t}. \label{D2} \end{equation}

\noindent Using Eq.~\ref{timeder} and Eq.~\ref{sub1}, one can show that

\begin{equation} \frac{d^2 d_A}{dt^2} = \left(\frac{H^2}{a^2}\right) \frac{d^2 d_A}{dz_r^2} + \left(\frac{H^2}{a} - \frac{1}{2aH} \frac{d H^2}{dt} \right) \frac {d d_A}{dz_r} \label{D2a}. \end{equation}

\noindent Combining the first derivative term in Eq.~\ref{D2a} with the one in Eq.~\ref{D2} and using Eq.~\ref{timeder}, we obtain for the geodesic deviation equation:

\begin{equation} \left(\frac{H^2}{2 a^4}\right) \frac{d^2 d_A}{dz_r^2} + \left(\frac{H^2}{a^3} - \frac{1}{4a^3H} \frac{d H^2}{dt} \right) \frac{d d_A}{dz_r} + \left( \frac{3 H_0^2  \Omega_m}{4 a^5} \right) d_A = 0 . \label{geod_dev2} \end{equation}

\noindent Finally, using the definition of $H^2$ in Eq.~\ref{hubble} to write $H^2$ in terms of the redshift $z_r$ (or equivalently $a$), simplification of each term and dividing through by $H_0^2/(2a^5)$ results in the Dyer-Roeder equation for the angular diameter distance as a function of the redshift:

\begin{equation} f_1(z_r) \frac{d^2 d_A}{dz_r^2} + f_2(z_r) \frac{d d_A}{dz_r} + f_3(z_r) d_A = 0 \label{dyer1}, \end{equation}

\noindent where

\begin{eqnarray} f_1(z_r) &=& (1+z_r)\left( 1+z_r\Omega_m-\Omega_\Lambda + \frac{\Omega_\Lambda}{(1+z_r)^2} \right)  \\
f_2(z_r) &=&  \frac{7}{2} z_r \Omega_m + \frac{1}{2} \Omega_m + 3 + \Omega_\Lambda\left(\frac{2}{(1+z_r)^2} - 3\right) \\ f_3(z_r) &=&  \frac{3}{2} \Omega_m . \label{dyer1funcs} \end{eqnarray}

\noindent Equation \ref{dyer1} agrees with the expressions for the Dyer Roeder equation in the book by Schneider et al, \cite{ehlers} section 4.5 when $\Omega_\Lambda$ is zero and with the expressions used by in more recent studies \cite{giovi}.

\subsection{Boundary Conditions and Scale}\label{boundary:sec}

The Dyer-Roeder equation, Eq.~\ref{dyer1}, is a second order ordinary differential equation for the angular diameter distance as a function of the redshift.  Because we are integrating backwards in time, or equivalently from the observer to the source, the bundles of light come to a point at the observer and the sources are extended and modeled by the tips of the geodesic deviation vectors.  In what follows, we will want to integrate a version of this equation numerically, so we need to find both the appropriate boundary conditions at $z_r=0$ and a suitable scaling for the numerical integration.

To find the boundary conditions, we first consider the geodesic deviation equation, Eq.~\ref{geod_dev}, in flat space where both the Ricci and Weyl curvature tensors are zero. Hence, we'd have

\begin{equation} D^2 \xi = 0, \label{geod_flat} \end{equation}

\noindent and we could take $\xi$ to be real valued.  Because $\ell^a$ is tangent to light rays, if we were working in cartesian coordinates, $\ell^a$ would have the form $\ell^a = (1/\sqrt{2}) (\dot t, \dot x, \dot y, \dot z)$, where the normalization factor of $1/\sqrt{2}$ is chosen to match the factor in our cosmological case, Eq.~\ref{ell1}.  Then, $D = \ell^a \partial_a = (1/\sqrt{2}) d/ds$, and Eq.~\ref{geod_flat} would be equivalent to

\[ D \xi = \frac{1}{\sqrt{2}}\frac{d \xi}{ds}  = c, \]

\noindent with solution $\xi = \sqrt 2 c s$ using the boundary condition that $\xi=0$ at the observer located at $s=0$ and with a constant $c$ to be determined.  To match this idea with the angular diameter distance, we think of some stick of length $L$ held at distance $s=s^*$, the tips of which subtend an angle $\alpha$ for an observer at $s=0$.  Then at $s^*$, identifying the length of the geodesic deviation vector with the stick of length $L$, we have $\xi=L = \sqrt 2 c s^* = \alpha s^*$ so that $c = \alpha/\sqrt 2$.  All of this implies that in flat space, to associate the solution of the geodesic deviation equation with length of a stick subtending an angle of $\alpha$ at an observer at $s=0$, we need $\xi \rightarrow 0$ and $D \xi \rightarrow \alpha/\sqrt 2$ as $s \rightarrow 0$.

In a cosmological space-time, in the local region of the observer, the cosmological effects are small, so that the boundary conditions we want to use for the solution to the Dyer-Roeder equation, Eq.~\ref{dyer1}, are the same as in flat space.  The difference is that we need to translate the action of the derivative operator $D$ into the action of the derivative operator $d/dz_r$.  Using

\[ D = \ell^0 \frac{\partial}{\partial t}, \]

\noindent with $\ell^0 = -1/(\sqrt{2} a)$ and $z_r = 1/a(t) -1$, and using the boundary condition for the first derivative of the geodesic deviation vector for flat space,  we have

\begin{equation} \lim_{z_r\rightarrow 0} D \xi = \lim_{z_r \rightarrow 0} \frac{1}{\sqrt 2} \frac{H}{a^2} \frac{d \xi}{d \, z_r} = \frac{H_0}{\sqrt 2} \left( \frac{d \xi}{d \,  z_r} \right)_{z_r=0} = \frac{\alpha}{\sqrt{2}}. \label{bc1} \end{equation}

\noindent  Transforming the geodesic deviation vector to the angular diameter distance, i.e., defining $d_A = \xi/\alpha$, implies that the boundary conditions we wish to use for the Dyer-Roeder equation, Eq.~\ref{dyer1} are

\begin{eqnarray} d_A (z_r = 0) &=& 0 \\ \left( \frac{d d_A}{d \,  z_r} \right)_{z_r=0} &=& \frac{1}{H_0}. \label{dyer_bc} \end{eqnarray}

Numerical solutions to differential equations are simpler if the scale of the differential equation is set so that the quantities of interest have roughly unit size.  The simplest way to accomplish this is to define an age of the universe $t_u$ by the relation

\begin{equation} a(t_u) = 1. \end{equation}

\noindent  For example, in a flat, $\Omega_\Lambda$-$\Omega_m$ space \cite{ryden}, the scale factor $a(t)$ is given by

\begin{equation} a(t) = \left( \frac{\Omega_m}{\Omega_\Lambda}
\right)^{1/3} \left\{ \sinh\left( \frac{3 H_0 \sqrt{\Omega_\Lambda} t}{2} \right) \right\}^{2/3}, \label{a1}  \end{equation}

\noindent and we can solve $a(t)=1$ for $t=t_u$ to obtain

\begin{equation} t_u =\frac{2}{3H_0} \frac{\sinh^{-1}( \sqrt{\Omega_\Lambda/\Omega_m})}{\sqrt{\Omega_\Lambda}}. \label{tu} \end{equation}

\noindent The age of the universe, $t_u$, provides a convenient numerical scale of $d_A$.  For our numerical work, we compute $\tilde d_A = d_A/t_u$.

The Dyer-Roeder equation, Eq.~\ref{dyer1}, does not explicitly show the functional form of the scale factor $a(t)$.  However, specifying the values of $\Omega_m$ and $\Omega_\Lambda$ simultaneously determines the numerical constants in the Dyer-Roeder equation and the particular form of the scale factor $a(t)$ for $k=-1,$ $0,$ or $+1$ FRW cosmological space-times.  In addition to finding the angular diameter distance using the Dyer-Roeder equation, one can use the line element, Eq.~\ref{FRW1} to compute the angular diameter distance.  To do so, first one specifies the final redshift of an emitter  and solves $1/a(t_e) = 1+z_r$ for $t_e$, the time (in the past) that the light was emitted.  Then, if the observer is at the origin, the integral

\[ \int_{t_e}^1 \, \frac{dt}{a(t)} =  d_r\]

\noindent determines the coordinate distance from the observer to the location of the emitter, with the metric distance given by $d_p = a^2 d_r$.  Then, the angular diameter distance is given by $d_A = (1+z_r)*d_p = d_r/(1+z_r)$.  One can confirm that the choice of boundary conditions given in this section provide the same angular diameter distance computed either from the Dyer-Roeder equation or from the metric distance.

\begin{figure}
\begin{center}
\scalebox{0.8}{\includegraphics{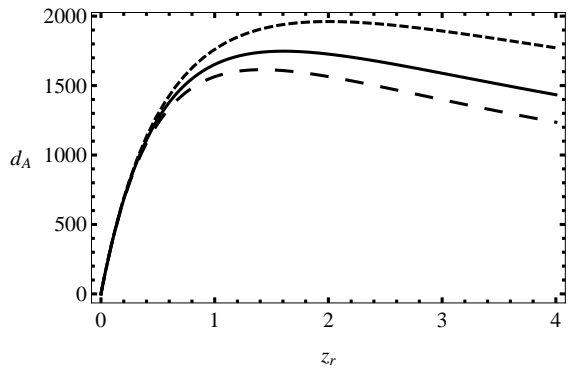}} \caption{\label{dyerplots}
Three plots of the angular diameter distance as a function of the redshift for different cosmological models as a function of redshift, $z_r$.  The angular diameter distance is measured in Mpc. The open cosmology with $\Omega_m=0.15$ and $\Omega_\Lambda = 0.7$ (short dashing) shows the highest peak, but also smallest decline in the angular diameter distance when compared with the favored flat model (solid line) of $\Omega_m=0.3$ and $\Omega_\Lambda = 0.7$ and a closed model (long dashing) with $\Omega_m=0.45$ and $\Omega_\Lambda = 0.7$.} \end{center}\end{figure}

Figure~\ref{dyerplots} shows the results of integrating the Dyer-Roeder equation, Eq.~\ref{dyer1}, for three choices of cosmology.  In each cosmology, we see that the angular diameter distance eventually begins to decline at high redshift due to the cosmic expansion.  Also, the choice of model impacts the maximum value of the angular diameter distance.

\section{Geodesic Deviation in Perturbed Cosmological Spaces} \label{perturbed:sec}

We are particularly interested in studying the Dyer-Roeder equation in space-times where a {\emph{flat}} FRW metric is perturbed by a gravitational lens. Thus, the metric of interest to us is

\begin{equation}  ds^2 =  - (1+2\varphi) dt^2 + a^2(t) (1-2\varphi) (dx^2 + dy^2 + dz^2). \label{metric:pert} \end{equation}

\noindent We will assume, as is now increasingly consistent with observational evidence, that the universe is a $\Lambda$ dominated CDM cosmology with $\Omega_\Lambda \approx 0.7$ and $\Omega_m \approx 0.3$.  We are particularly interested in the Dyer-Roeder equation in the context of gravitational lensing, so that we will take the  gravitational perturbation $\varphi$ as representing one or more galaxy clusters at medium redshift along the line of sight from the observer to some distant, observed galaxies at higher redshift.

Again, we will examine the geodesic deviation equation in the form of Eq.~\ref{geod_dev}, so we need to compute the NP components $\Phi_{00}$ and $\Psi_0$ of the Ricci and Weyl tensors.  Since the metric is first order in the gravitational potential, we will compute the Ricci and Weyl tensor components to that order.  In addition, terms of order $\varphi \times \dot a /a$ or further derivatives of these terms are small and will be considered higher order terms.  To lowest order the non-zero Ricci tensor components are

\begin{eqnarray} R_{00} &=& -\frac{3 \ddot a}{a} + \frac{1}{a^2} \nabla^2 \varphi  \nonumber \\ R_{ii} &=& 2 \dot a^2 + a\ddot a + \nabla^2 \varphi, \label{ricci:comp} \end{eqnarray}

\noindent where $i$ represents any of the spatial components.  We see that the Ricci tensor components for perturbed metric are consistent with the Ricci tensor components of the unperturbed metric in Eq.~\ref{Ricci:general}.  The Weyl tensor is no longer zero, and only reflects contributions from the perturbation.  To lowest order in $\varphi$, we obtain

\begin{eqnarray}  C_{0i0i} =  \varphi_{ii} - \frac{1}{3} \nabla^2 \varphi &\quad\quad& C_{0i0j} = \varphi_{ij} \quad i\ne j \nonumber \\ C_{ijij} = -a^2 (\varphi_{kk} - \frac{1}{3} \nabla^2 \varphi) &\quad\quad& C_{ijik}  = a^2 \varphi_{jk} \quad i \ne j \ne k, \label{wely:comp} \end{eqnarray}

\noindent where the subscript $i$'s on the potential denote derivatives with respect to a spatial coordinate.  The Weyl and Ricci tensors of the perturbation are consistent with the versions found in previous work under for perturbations of flat metrics by weak gravitational fields \cite{kk}.

As our physical situation, we consider the gravitational perturbation $\varphi$ to be localized near the origin, an observer to be situated along the $+\hat z$ axis, observing galaxies in the background of the perturbation, so that the perturbation is acting as a gravitational lens.  We will take complex stereographic coordinates $\zeta$ to span the celestial sphere of the observer, where $\zeta=0$ corresponds to the spatial part of past-directed light rays moving along the $\hat z$ axis towards the origin.  Since we are interested in examining the Dyer-Roeder equation in the context of gravitational lensing, a small angle approximation can be applied so that we are only interested in the rays for which $\zeta$ is small.  Further, we are interested in computing the $\Phi_{00}$ and $\Psi_0$ NP Ricci and Weyl tensor components in to first order in the perturbation $\varphi$ taking into account the slow change in the scale function $a(t)$ during the epoch for gravitational lensing.  To accomplish this, we consider the $\ell^a$ and $m^a$ tetrad vectors to be null vectors in the background FRW metric, not the perturbed metric, and products of $\zeta$ with $\varphi$ or $\dot a/a$ are higher order terms that we will discard.

In Cartesian coordinates, the tangent vector to the past directed light rays of the background FRW metric is given by

\begin{equation} \ell^a = \frac{1}{\sqrt{2} a (1+\zeta\bar\zeta)} \left( -1 - \zeta\bar\zeta , \frac{\zeta + \bar\zeta}{a}, \frac{i(\bar\zeta - \zeta)}{a}, \frac{-1+\zeta\bar\zeta}{a} \right). \label{ella} \end{equation}

\noindent The complex, spatial vector in the null tetrad is

\begin{equation} m^a = \frac{1}{\sqrt{2} (1+\zeta\bar\zeta)} \left(0, \frac{1-\bar\zeta^2}{a}, -i\frac{1+\bar\zeta^2}{a}, \frac{2\bar\zeta}{a} \right). \label{ma} \end{equation}

\noindent One can check that $\ell^a$ and $m^a$ satisfy $g_{ab} \ell^a\ell^b=g_{ab} m^a m^b= g_{ab} \ell^a m^b=0$ and $g_{ab} m^a \overline m^b = 1$ in the background ($\varphi=0$) FRW metric, and that $\ell^a \nabla_a \ell^b = \ell^a \nabla_a m^b =0$ as well, using the connection associated with the background metric.

Then to first order in the perturbation, the NP Ricci tensor component we need is

\begin{equation} \Phi_{00} =\frac{1}{2a^2} \left( \frac{\dot a^2}{a^2} - \frac{\ddot a}{a}  + \frac{1}{a^2} \nabla^2 \varphi \right) = \frac{3}{4}\frac{H_0^2 \Omega_m}{a^5} + \frac{1}{2a^4} \nabla^2 \varphi, \label{Ricci:pert} \end{equation}

\noindent which we see has split into two terms: one for the background FRW metric and one from the perturbation.  For convenience, we chose to write

\begin{equation} \Phi_{00} = \frac{3}{4}\frac{H_0^2 \Omega_m}{a^5} + \Phi_{00p}, \end{equation}

\noindent where the $p$ subscript on the latter part of $\Phi_{00}$ reminds us that this part of the Ricci tensor is due to the perturbation.  The Weyl tensor component is given by

\begin{equation} \Psi_0 = \frac{1}{2a^4} \left( \varphi_{xx} - \varphi_{yy} - 2i\varphi_{xy} \right), \label{weyl:pert} \end{equation}

\noindent where the subscripts on the gravitational potential indicate derivatives with respect to the coordinates $x$ or $y$.  Since the Weyl tensor is zero in non-perturbed FRW metrics, we can think of all of the Weyl tensor as being due to the perturbation.

Because the perturbation breaks the homogeneity of the background metric, we can no longer assume that all the components of the geodesic deviation vector $q^a$ vary in the same manner along the central light ray.  We assume that the geodesic deviation vector takes the form

\[ q^a = \left(0, \frac{q^x}{a}, \frac{q^y}{a}, \frac{q^z}{a} \right) \]

\noindent where the $q^i$ components are small.  The product of $q^i$ with $\zeta$ will be higher order, so that we assume that the basis vector $m^a$ takes the form

\[ m^a \approx \left( 0, \frac{1}{\sqrt 2 a}, \frac{ -i}{\sqrt 2 a}, 0 \right), \]

\noindent and the complex components of $\mathbf{z}$ are given by

\begin{equation} \xi = g_{ab} q^a m^b = \frac{1}{\sqrt 2} (q^x - i q^y) \label{q:comp} \end{equation}

\noindent to lowest order.  The geodesic deviation equation, Eq.~\ref{geod_dev}, separates into two real, coupled ordinary differential equations as

\begin{eqnarray} D^2 q^x &=& -\Phi_{00} q^x - \Psi_{0r} q^x + \Psi_{0i} q^y \nonumber \\ D^2 q^y & = & -\Phi_{00} q^y + \Psi_{0r} q^y + \Psi_{0i} q^x \label{pert:geod_dev} \end{eqnarray}

\noindent where we have separated the real and imaginary parts of the Weyl tensor as $\Psi_0 = \Psi_{0r} + i \Psi_{0i}$.

We now wish to transform the geodesic deviation equation, Eq.~\ref{pert:geod_dev}, by changing the derivative from the directional derivative $D$ to a derivative with respect to the redshift, $z_r$, and also by separating the part of the curvature, $\Phi_{00}$ due to the background FRW metric from the perturbation.  This gives us an analog of the Dyer-Roeder equation with an explicit source term due to the perturbation, where the algebra leads to the same functions,  $(f_1, f_2, f_3)$ as in Eq.~\ref{dyer1funcs}:

\begin{eqnarray}   f_1(z_r) \frac{d^2 q^x}{dz_r^2}  +   f_2(z_r) \frac{d q^x}{dz_r} +  f_3(z_r)  q^x & = &  \left(\frac{2a^5}{H_0^2} \right) \left( - \Phi_{00p} \, q^x -  \Psi_{0r} \, q^x + \Psi_{0i} \, q^y \right) \\
 f_1(z_r) \frac{d^2 q^y}{dz_r^2} +  f_2(z_r) \frac{d q^y}{dz_r} +  f_3(z_r)  q^y   &=&  \left(\frac{2a^5}{H_0^2} \right) \left( - \Phi_{00p} \, q^y + \Psi_{0r} \, q^y + \Psi_{0i} \, q^x  \right) . \label{dyer:pert} \end{eqnarray}

\noindent The appearance of the term $2a^5/H_0^2$ on the right hand side comes from the re-scaling from Eq.~\ref{geod_dev2}.

\section{Angular Diameter Distances for an Axially Symmetric Thin Lens}\label{lensdyer:sec}

In the case that the gravitational perturbation of the background FRW space-time is a single lens axially symmetric along an axis connecting the observer to the center of the lens, one knows, a priori, that there are certain directions in which a geodesic deviation vector can point so that it may grow or shrink, but will not rotate.  For example, if the lens is symmetric about the origin, a light ray traveling to an observer located on the $+\hat z$ axis in the $\hat x$-$\hat z$ plane will have two geodesic deviation vectors, one that points in the $+\hat x$ direction and another that points in the $+\hat y$ direction that do not rotate.  This can be seen because for an axial symmetric lens, for a ray in the $\hat x$-$\hat z$ plane, the derivative $\varphi_{xy}$ is zero along the entire light ray, so that $\Psi_{0i}=0$.  Thus, the geodesic deviation equations for the two components,  $(q^x, \, q^y)$, of the geodesic deviation vector separate, and we can consider two independent vectors, one where $q^y_{1}=0$ and one where $q^x_{2}=0$.

In the FRW background metric, one can simply assume that $q^x$ is the length of the stick whose tips are subtended by the angle $\alpha$, so that $d_A = q^x/\alpha$.  In the case of the axially symmetric lens, we assume that we are tracing a pencil of light rays backwards in time from the observer to a redshift $z_s$ of sources beyond the redshift of the lens $z_l$, so that $z_s>z_l$.  The general situation calls for a circular shaped source so that we can take $q^x_{1} = q^y_{2} = t \ll 1$ at the source galaxy redshift, $z_s$.  At the observer at $z=0$, the boundary conditions would be that $q^x_{1} = q^y_{2} = 0$ but that derivatives would be given by

\[ \left(\frac{d q^x_{1}}{dz_r}\right)_{z_r=0}  = \frac{\alpha}{H_0}, \quad\quad \left(\frac{d q^y_{2}}{dz_r}\right)_{z_r = 0} = \frac{\beta}{H_0},\]

\noindent where the angles $\alpha$ and $\beta$ represent the opening of the observed ellipse with $\beta>\alpha$ for a standard axially symmetric gravitational lens.  The angular diameter distance would be given by $\sqrt{q^x_{1}\,q^y_{2}/(\alpha \beta)}$.  By doing ray shooting, one can find the angles $\alpha$ and $\beta$ to match the condition $q^x_{1} = q^y_{2} = t$.

However, it is simpler to proceed by assuming that at the observer, the pencil of rays converges equally in all directions, or that the pencil has a circular cross-section at the observer opening at an angle $\alpha$.  At the source, the initially circular pencil of rays will have formed a small ellipse, and if the light ray travels in the $\hat x$-$\hat z$ plane, the vector $q^a_{1}$ will form the semi-major axis parallel to the $\hat x$ direction and $q^a_2$ will form the semi-minor axis parallel to the $\hat y$ direction.  (The source which appears as a circle at the observer must be an ellipse oriented opposite the usual observed elliptical image from weak gravitational lensing.)  The area associated with this ellipse is proportional to $q^x_{1} q^y_{2}$, and the angular diameter distance will be $d_A = \sqrt{q^x_{1} q^y_{2} / \alpha^2}$ where $q^x_1$ and $q^y_2$ take their values at the source galaxy redshift.

This implies that we should define scaled quantities $\delta_x = q^x_{1}/\alpha$ and $\delta_y = q^y_{2}/\alpha$. Following Section \ref{boundary:sec}, we will assume as our boundary conditions at the observer:

\begin{eqnarray} \delta_x (z_r=0) & = & \delta_y (z_r=0)  =   0 \\
 \left(\frac{d\delta_x}{d\, z_r} \right)_{z_r=0} & =&   \left(\frac{d\delta_y}{d\, z_r} \right)_{z_r=0} = \frac{1}{H_0} \label{bc:pert} \end{eqnarray}

\noindent and the angular diameter distance is simply $\sqrt{\delta_x \delta_y}$.  The two equations for the two components of the geodesic deviation equation, written in terms of derivatives with respect to the redshifts, Eqs.~\ref{dyer:pert}, then become in this context two decoupled differential equations for $\delta_x$ and $\delta_y$.  It is convenient to divide each term by the function $f_1(z_r)$, so that our resulting equations for $\delta_x$ and $\delta_y$ take the form:

\begin{eqnarray}  \frac{d^2 \delta_x}{dz_r^2}  +    \frac{f_2(z_r)}{f_1(z_r)} \frac{d \delta_x}{dz_r}   +  \frac{f_3(z_r)}{f_1(z_r)} \delta_x  & =& - \left(\frac{2a^5}{H_0^2} \right) \left( \frac{ \Phi_{00p} + \Psi_{0r} } {f_1(z_r)} \right) \delta_x  \nonumber \\
 \frac{d^2 \delta_y}{dz_r^2} + \frac{f_2(z_r)}{f_1(z_r)} \frac{d \delta_y}{dz_r} + \frac{f_3(z_r)}{f_1(z_r)} \delta_y   &=& - \left(\frac{2a^5}{H_0^2} \right) \left( \frac{\Phi_{00p} - \Psi_{0r}}{f_1(z_r)} \right) \delta_y. \label{dyer:thinaxial} \end{eqnarray}

\noindent In a flat cosmology, with $\Omega_m + \Omega_\Lambda = 1$, there are no values of the redshift where $f_1(z_r) = 0$.  We will refer to Eqs.~\ref{dyer:thinaxial} as the perturbed Dyer-Roeder equation.

The perturbations on the right hand side of Eqs.~\ref{dyer:thinaxial} are continuous functions of the redshift that are evaluated along the light ray from the observer backwards in time to the source.  In principle, to evaluate the perturbed Dyer-Roeder equation one would need to simultaneously solve the null geodesic equations of the metric, Eq.~\ref{metric:pert} to evaluate the Ricci and Weyl tensor terms as one integrated our perturbed Dyer-Roeder equation.  If one wanted to examine thick gravitational lenses, one could integrate along the trajectory of the central light rays to examine the accuracy of strong gravitational lensing by truncated dark matter halos, re-parameterizing their null geodesic equations in terms of the redshift, and simultaneously find numerical solutions to the perturbed Dyer-Roeder equation.

It is much more common in gravitational lensing to consider the gravitational perturbation as thin lens, because the width of the lens's dark matter halo is small compared to the distance between the observer and the lens.  In this circumstance, one can approximate the lens as having a delta function form.  If we assume that the lens is a static perturbation centered at the origin in $(x,y,z)$ coordinates, we would write

\begin{eqnarray} \Psi_0 &=& ~_L\Psi_0 (x,y) \delta(z_r - z_l) \nonumber \\ \Phi_{00p} &=& ~_L\Phi_{00p} (x,y) \delta(z_r - z_l), \label{proj:pert} \end{eqnarray}

\noindent where the delta functions are written in terms of redshifts and place the lens location at the mean lens redshift $z_l$. Also, the scale factors, $a(z_r)$, in the perturbations are evaluated at the lens redshift.  The projected perturbations, denoted by the prescript $L$ for lens plane, are found by integrating the three dimensional Ricci and Weyl tensor terms along the $\hat z$ axis

\begin{eqnarray} ~_L\Psi_0 (x,y) &=& \int_{-\infty}^{+\infty} \, \Psi_0 (x,y,z) \, dz \\  ~_L\Phi_{00p} (x,y) &=& \int_{-\infty}^{+\infty} \, \Phi_{00p} (x,y,z) \, dz \label{proj}. \end{eqnarray}

\noindent The $(x,y)$ coordinates remaining in the projected perturbations become the coordinates in the lens plane through which our light ray passes.

One can show that the projected Weyl tensor perturbation, $_L \Psi_0$, is in fact the measured weak gravitational shear of the lens, and that the projected Ricci tensor perturbation, $_L \Phi_{00}$, is the projected mass density of the lens \cite{kk}.  Integrating the gravitational perturbation $\varphi$ in the metric, Eq.~\ref{metric:pert}, defines the projected gravitational potential used in lensing:

\begin{equation} \psi (x,y) = \int_{-\infty}^{+\infty} \, \varphi (x,y,z) \, dz. \end{equation}

\noindent The gravitational lensing shear and projected mass densities are defined through second derivatives of the projected potential

\begin{equation} \gamma_1  = \frac{1}{2} \left( \psi_{xx} - \psi_{yy} \right) \quad\quad \gamma_2 = \psi_{xy}, \end{equation}

\begin{equation} \kappa = \frac{1}{2} \left( \psi_{xx} + \psi_{yy} \right). \end{equation}

\noindent Then, from Eqs.~\ref{Ricci:pert} and \ref{weyl:pert}, the projection of the perturbed Ricci and Weyl tensor terms into a lens plane clearly allows one to identify these perturbations as the weak lensing mass density and shear.

The projected perbutations in Eqs.~\ref{proj:pert} are related by a second order, partial differential equation in the form of Poisson's equation, where derivatives of the projected Weyl tensor (which one can measure directly) act as a source term for the projected Ricci tensor as derived in \cite{kc} or \cite{ss} in a different way.  This means that in the context of weak gravitational lensing studies, one can measure and determine the perturbations that act as source terms in Eqs~\ref{dyer:thinaxial}, and then one can use the perturbed Dyer-Roeder equation to determine the variation of the angular diameter distances of sources behind the lens plane.  The thin-lens plane version of the perturbed Dyer-Roeder equation is then

\begin{eqnarray}  \frac{d^2 \delta_x}{dz_r^2}  +    \frac{f_2(z_r)}{f_1(z_r)} \frac{d \delta_x}{dz_r}   +  \frac{f_3(z_r)}{f_1(z_r)} \delta_x  & = &  \nonumber \\ &- & \left(\frac{2a^5 \delta_x \, (_L\Phi_{00p} + _L\Psi_{0r})}{H_0^2 \, f_1(z_r) } \right)  \delta(z_r-z_l) \nonumber \\
 \frac{d^2 \delta_y}{dz_r^2} + \frac{f_2(z_r)}{f_1(z_r)} \frac{d \delta_y}{dz_r} + \frac{f_3(z_r)}{f_1(z_r)} \delta_y   &=& \nonumber \\ & - & \left(\frac{2a^5 \delta_y \,  (_L\Phi_{00p} - _L\Psi_{0r}) }{H_0^2 \, f_1(z_r)} \right) \delta(z_r-z_l) .   \label{dyer:thinaxial1} \end{eqnarray}

To integrate this version of the perturbed Dyer-Roeder equation, where the perturbation is taking the form of a delta function at the redshift of the lens plane, one simply needs to integrate the homogeneous ordinary differential equation with zero perturbation from the observer at $z_r=0$ up to the lens plane, reset the boundary conditions, and restart the integration of the homogeneous equation.  At the observer, we will use the boundary conditions as in Eq.~\ref{bc:pert}, and integrate until reaching the lens plane at $z_r=z_l$.  At the lens plane, the boundary conditions we want are that $\delta_x$ and $\delta_y$ are continuous, but that the first derivatives on each side of the lens plane are discontinuous according to

\begin{eqnarray} \left( \frac{d \delta_x}{dz_r} \right)_{z_l^+} & = & \left( \frac{d \delta_x}{dz_r} \right)_{z_l^-} -  \left(\frac{2a^5}{H_0^2} \right) \left( \frac{ _L\Phi_{00p} + _L\Psi_{0r} } {f_1(z_l)} \right) ~_L\delta_x  \nonumber \\   \left( \frac{d \delta_y}{dz_r} \right)_{z_l^+} & = & \left( \frac{d \delta_y}{dz_r} \right)_{z_l^-} - \left(\frac{2a^5}{H_0^2} \right) \left( \frac{ _L\Phi_{00p} - _L\Psi_{0r} } {f_1(z_l)} \right) ~_L\delta_y, \label{bclens} \end{eqnarray}

\noindent where $_L \delta_x$ and $_L \delta_y$ are the values of $\delta_x$ and $\delta_y$ at the lens redshift.  If one is doing a numerical integration, Eq.~\ref{bclens} tells us how to set the value of the first derivative after reaching the lens plane, or at $z_l^+$ based on the value at reached in integrating the homogeneous equation up to the lens plane at $z_l^-$.  On can derive these boundary conditions by integrating the thin lens version of the perturbed Dyer Roeder equation, Eq.~\ref{dyer:thinaxial1}, from $z_l-\epsilon$ to $z_l+\epsilon$ and taking the limit $\epsilon \rightarrow 0$, with the assumption that all the functions are continuous at the lens plane location.

\section{Truncated Navarro, Frenk and White Model}

As an example, we consider a truncated Navarro Frenk and White (tNFW) model, introduced by Baltz et al. \cite{baltz}, whose three dimension matter potential is given by

\begin{equation} \rho(r_p) = \frac{\delta_c \rho_{\mbox{\scriptsize crit}}}
{\left(\frac{r_p}{r_s} \right) \left( 1 + \frac{r_p}{r_s} \right)^2
\left(1 + \left(\frac{r_p}{r_t} \right)^2\right)}.
\label{baltz:rho}\end{equation}

\noindent where $r_p = a_l r$ is a proper radius at the lens redshift and $a_l = 1/(1+z_l)$, $\rho_{\mbox{\scriptsize crit}}$ is the critical density at the lens redshift, $r_s$ is a scale radius defined as the peak of $r^2\rho(r)$, and $\delta_c$ is a characteristic density contrast.  The characteristic density contrast is related to a concentration parameter, $c$ by

\[ \delta_c = \frac{200}{3}\frac{c^3}{\log{(1+c)}-c/(1+c)}. \]

\noindent The tidal radius $r_t$ is a radius inside which the tNFW model well approximates the standard NFW model.  The truncation is introduced so that the matter density has a finite total mass when integrated over all space.  The gravitational potential associated with this matter density is given by

\begin{eqnarray}
\varphi(x_p)  =  \frac{GM_0}{r_s}\frac{\tau^2}{(1+\tau^2)^2} & \Bigg[ &
 \frac{\pi(\tau^2-1)}{2\tau}
-2\ln\tau + \arctan (x_p/\tau) \Big( \frac{1}{\tau}-\tau-2\frac{\tau}{x_p} \Big) \nonumber \\ ~ &~ &
+\ln\left(\frac{1+(x_p/\tau)^2}{(1+x_p)^2}\right)\Big(
\frac{\tau^2-1}{2x_p} -1\Big) \Bigg] \label{baltz:phi1}
\end{eqnarray}

\noindent with $x\equiv a_l r/r_s$. Here  $M_0 \equiv 4 \pi r_s^3 \delta_c\rho_{\mbox{\scriptsize crit}}$ and $\tau = r_t/r_s$.  For our numerical work, we use $r_s = 250$~kpc, $c = 6.593$, and set $\tau = 3c$ which creates good agreement between the NFW model and the tNFW model within the virial radius \cite{baltz}.  We place the lens at $z_l = 0.45$.  The total mass for this parameter choice is $1.5 \times 10^{15} M_\odot$.

\begin{figure}
\begin{center}
\scalebox{0.8}{\includegraphics{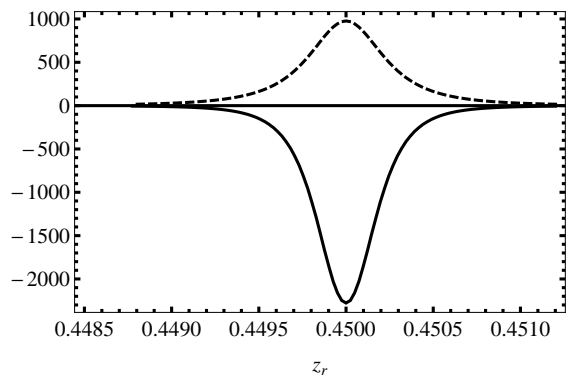}} \caption{\label{ricciweyl:fig}
The Newman-Penrose formalism components of the Ricci (dashed) and Weyl (solid) tensor components in dimensionless units as a function of the redshift for a truncated NFW potential whose center is at redshift $z_l=0.45$.  The potential is chosen so that the overall mass of the perturbation is $1.5 \times 10^{15}\, M_\odot$.  The values represented here are along a ray with a constant $x$ parameter value where $x = D_l \theta / a_l$ where $D_l$ is the angular diameter distance to the lens.  For this plot, $\theta = 120$~arc sec.  The placement of the $a_l=(1+z_l)^{-1}$ in the denominator makes $x$ a coordinate value and $a_l x$ a physical distance on which the potential and NP tensors depend.} \end{center}\end{figure}

To project the Ricci and Weyl tensor components into the lens plane, we need to integrate along the $\hat z$ axis as in Eq.~\ref{proj}.  Because closed form expressions of $_L \Phi_{00p}$ and $_L \Psi_0$ are cumbersome, it is simpler to perform a numerical integration.  Figure~\ref{ricciweyl:fig} shows the dimensionless Ricci and Weyl tensor perturbations as a function of the redshift in the region near the lens at $z_l=0.45$ for a ray in that strikes the lens plane along the $\hat x$ axis at an angle of $\theta = 120$~arc sec.

\begin{figure}
\begin{center}
\scalebox{0.8}{\includegraphics{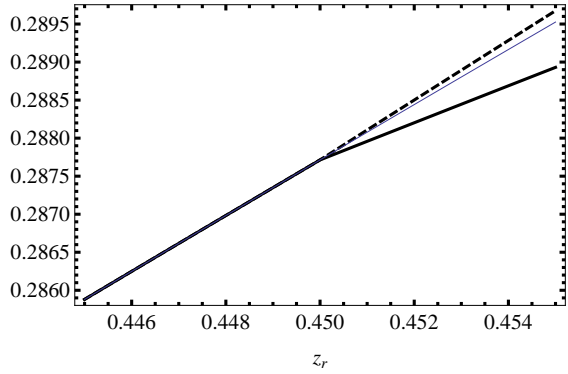}} \caption{\label{kink:fig}
The dimensionless, scaled geodesic deviation vectors $\delta_x$ (dashed) and $\delta_y$ (heavy solid) as a function of the redshift near the lens for a ray making an angle of $120$ arc sec with the central axis. We see that both components are continuous at the lens plane, with different discontinuities in the first derivatives caused by the source terms in on the right hand side of Eq.~\ref{dyer:thinaxial1}.  The thin line corresponds to the solution to the unperturbed Dyer-Roeder equation for reference.} \end{center}\end{figure}

\begin{figure}
\begin{center}
\scalebox{0.8}{\includegraphics{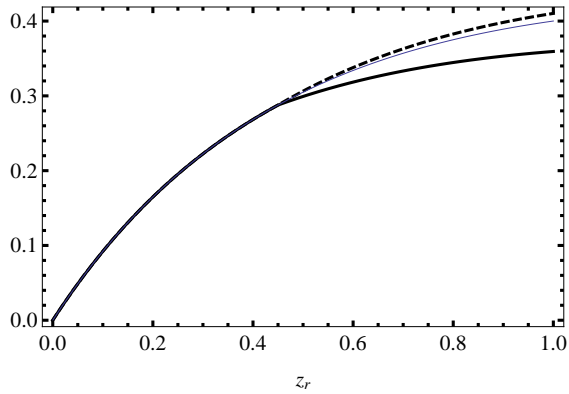}} \caption{\label{deltas:fig}
The dimensionless, scaled geodesic deviation vectors $\delta_x$ (dashed) and $\delta_y$ (heavy solid) as a function of the redshift for a ray making an angle of $120$ arc sec with the central axis.  The thin solid like represents the integration of the unperturbed Dyer-Roeder equation, so that we see that at large angles, the $x$ component increases and $y$ component decreases relative to the unperturbed case, turning a cross-sectional bundle that was circular at the observer into an ellipse extended along the $\hat x$ direction.} \end{center}\end{figure}

In Fig.~\ref{kink:fig}, we show the continuity of the $\delta_x$ and $\delta_y$ functions and their discontinuous first derivatives at the lens plane for a ray making an angle of $120$ arc sec with respect to the axis of symmetry as a function of the redshift.  We see that for a ray at this radius, the $\delta_x$ component represented by the dashed line is increased and the $\delta_y$ component is decreased relative to the values they would have if there was no lens (given by the light line).  This is consistent with the appearance of a circular cross section at the observer.  Figure~\ref{deltas:fig} shows the two components over a wider range of redshift.  Figs.~\ref{kink:fig} and \ref{deltas:fig} show that the ellipse formed by the action of the lens on a circular bundle of rays emitted from the observer will have its semi-major axis orient along the axis parallel to the line joining the center of the mass distribution with the point where the light ray pierces the lens plane.

\begin{figure}
\begin{center}
\scalebox{0.8}{\includegraphics{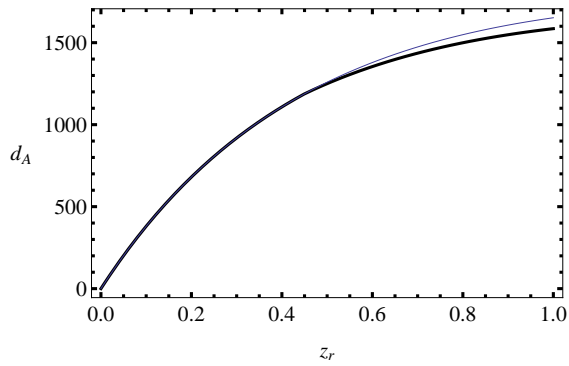}} \caption{\label{distance:fig}
The angular diameter distance (heavy line) in Mpc as a function of redshift for an object making an angle of $120$ arc sec from the center of a truncated NFW model whose total mass is $1.5 \times 10^{15} M_\odot$.  The light solid line represents the angular diameter distance as function of redshift in the absence of a lens.  The effect of the focusing theorem is seen in that the angular diameter distances match until reaching the perturbation of the gravitational lens.  Thereafter, the true angular diameter distance is less than the angular diameter distance if there was no lens present.} \end{center}\end{figure}

Figure~\ref{distance:fig} shows the angular diameter distance in Mpc as a function of redshift for objects behind a gravitational lens at $z_l=0.45$ when the object makes an angle of $120$ arc sec with respect to the axis of symmetry for our toy model of a truncated NFW potential whose total mass is $1.5\times 10^{15} M_\odot$.  Here, we see a confirmation of the focussing theorem which states that the angular diameter distance to distance source is smaller in the presence of a perturbation due to the focusing of the light \cite{seitz}, \cite{ehlers}.

\begin{figure} \begin{center} \scalebox{0.8}{\includegraphics{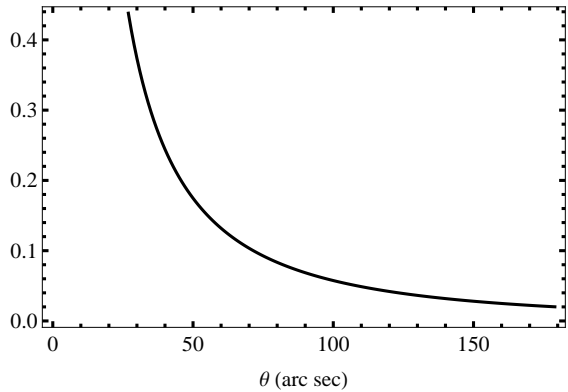}} \caption{\label{angdiamwide:fig} The relative error in the angular diameter distance for sources at a redshift of 1.0 given a truncated NFW lens at redshift of 0.45 with a total mass of $1.5 \times 10^{15} M_\odot$. Ignoring the perturbation of the lens can impact the determination of the angular diameter distance significantly for observation angles less than $100$ arc sec.  Even at $180$ arc sec, ignoring the perturbation of the lens leads to over-estimating the angular diameter distance by over $2\%$.} \end{center}\end{figure}

Finally, in Fig.~\ref{angdiamwide:fig} we show the relative difference between true angular diameter distance taking into account the perturbation and the angular diameter distance without the perturbation as a function of the distance from the center of the lens at a final source redshift of $z_r=1$.  At small observation angles near the Einstein ring radius, the unperturbed Dyer-Roeder equation's solution over-estimates the angular diameter distance by nearly $30\%$.  Even at relatively large observation angles, say $180$ arc sec, the unperturbed angular diameter distance is $2\%$ larger than the correct value taking into account the perturbation of the lens.

In a number of papers, for instance \cite{kaiser}, an average angular diameter distance for all sources back to some redshift is studied.  It has been found that the average error is about $1\%$ across the sky.  Modeling the error curve in Fig.~\ref{angdiamwide:fig} as an inverse power series

\[ f(r) = a_0 + \frac{a_1}{r} + \frac{a_2}{r^2} \]

\noindent allows one to fit the error curve very accurately.  To first order, we can assume that the density of objects observed behind the lens is constant as a function of radial distance from the center of the lens.  We can then compute an average difference between the perturbed angular diameter distance and unperturbed angular diameter distance across the sky.  Our result for this model is $6.5\%$ averaging all the objects that would be observed in an annulus between $30$ and $180$ arc sec.  This annulus is the region where one would typically identify statistical weak lensing objects; at less than $30$ arc sec, background objects are difficult to distinguish from cluster members in ground-based observations.  The $6.5\%$ value reflects the contribution from the larger number of objects at higher radius that have lower differences in angular diameter distance.  An averaging across the entire sky would need to model the frequency of lens objects, and the lens studied here would be relatively rare.  Thus, the difference between the perturbed and unperturbed angular diameter distances shown in Fig.~\ref{angdiamwide:fig} appear consistent with the literature.

\section{Discussion}

The modern use of the Newman-Penrose spin coefficient formalism is typically in vacuum space-times where the Ricci tensor is zero and the interest is in determining properties of the gravitational radiation coded into components of the Weyl tensor.  To our knowledge, this paper is the first attempt to understand how the NP formalism would be applied to a perturbed cosmological space-time. We show that the geodesic deviation equation in the NP formalism for a non-perturbed FRW metric very directly leads to the Dyer-Roeder equation for the angular diameter distance as a function of the redshift.

Of particular interest to us is the application of this technique for perturbations of a flat FRW metric by a gravitational lens.  In previous work \cite{kk}, it has been shown that the projection of the $\Phi_{00}$ Ricci tensor component is the gravitational lensing convergence, frequently denoted by $\kappa$, and the projection of the $\Psi_0$ Weyl tensor component is the gravitational lensing shear, written as $\gamma_1 + i \gamma_2$.  In practice, one measures $\gamma_1$ and $\gamma_2$ directly from the observation of distance galaxies behind the gravitational lens in a weak lensing measurement, and from that infers the $\kappa$ value.  This leads to our motivation for studying the geodesic deviation equation in the NP formalism.  Given that in gravitational lensing one needs to find angular diameter distances to set the scale of the measurement, a way to systematically and easily use the measured perturbation to the space-time metric in an equation for the angular diameter distance is of significant value.  The derivation of the Dyer-Roeder equation, and the perturbed Dyer-Roeder equation, presented here using the Newman-Penrose formalism, accomplishes this objective.

The presentation of a perturbed Dyer-Roeder equation in this paper is different from those used \cite{clarkson1} or similar papers that use the optical scalars to mediate the influence of the gravitational lens.  These papers relate the weak lensing observables to the optical scalars and generate a set of first order differential equations for the angular diameter distance.  Our approach utilizes the weak lensing observables directly in second order differential equations with the lensing observables acting as the source terms.  We believe our equivalent approach is conceptually simpler and easier to apply, particularly if more than one lensing event occurs along a given geodesic.

In a simple test model, we see that the impact of ignoring the effect of the perturbation on the angular diameter distance to sources behind the gravitational lens can be large along individual geodesics.  The average for objects behind a large gravitational lenses is about $6.5\%$.  This result is consistent with those that average over the entire sky, but points out that significant variations in background distance measures will occur directly behind large gravitational lenses.

At the current time, one typically does not have particularly accurate measurements of the redshifts of the background sources used in gravitational lensing studies.  Therefore, in current studies, the effect of ignoring the perturbation of the lens in computing the approximate angular diameter distance to the average source galaxy may continue to be negligible.  However, in the long run, this source of error will become more significant as it becomes easier to collect light from faint background galaxies in multiple filters allowing for reasonable measurements of photometric redshifts.  At that future point, a method such as the perturbed Dyer-Roeder equation outlined in this paper will become useful in determining quickly the angular diameter distances to source galaxies.

\begin{acknowledgements}
AA thanks the Bridgewater State University Adrian Tinsley Program for
Undergraduate Research for a Summer Grant that enabled his
participation in this project.  Both authors would like to thank and recognize Ezra T. Newman, whose eureka moment prompted a deeper examination of this topic.
\end{acknowledgements}



\end{document}